\documentclass[twocolumn,prd,superscriptaddress,preprintnumbers,amsmath,amssymb,nofootinbib]{revtex4}
\usepackage{amssymb}
\usepackage{slashed}
\usepackage{graphicx}
\usepackage{graphics}
\usepackage{epsfig}
\usepackage{color}
\usepackage{soul}
\usepackage{tabularx}
\usepackage{amsmath}
\usepackage{amsfonts}
\usepackage{amssymb}
\usepackage{hyperref}
\usepackage{microtype}
\usepackage[normalem]{ulem}
\usepackage[utf8]{inputenc}

\newcommand{\be}{\begin{equation}}
\newcommand{\ee}{\end{equation}}
\newcommand{\bea}{\begin{eqnarray}}
\newcommand{\eea}{\end{eqnarray}}

\newcommand{\MWGC}{\ensuremath{M_{\text{\tiny WGC}}}}
\newcommand{\mWGC}{\ensuremath{m_{\text{\tiny WGC}}}}
\newcommand{\mWGCast}{\ensuremath{m_{\text{\tiny WGC}\,\ast}}}

\newcommand{\MPl}{\ensuremath{M_{\text{\tiny Pl}}}}

\newcommand{\eUone}{\ensuremath{e}}

\newcommand{\UniG}{\ensuremath{\rm U}}

\usepackage{color}
\usepackage{xcolor}
\usepackage{slashed}
\definecolor{mygreen}{rgb}{0, 0.6, 0}
\definecolor{orange}{rgb}{1, 0.5, 0}
\usepackage{empheq}

\begin{document}

\author{Senarath de Alwis}
\affiliation{Physics Dept. UCB390 University of Colorado, Boulder, CO 80309, U.S.A.}

\author{Astrid Eichhorn} 
\affiliation{CP3-Origins, University of Southern Denmark, Campusvej
  55, DK-5230 Odense M, Denmark} \affiliation{Institut f\"{u}r
  Theoretische Physik, Universit\"{a}t Heidelberg, Philosophenweg 16,
  69120 Heidelberg, Germany}

\author{Aaron Held}
\affiliation{Institut
 f\"{u}r Theoretische Physik, Universit\"{a}t Heidelberg,
 Philosophenweg 16, 69120 Heidelberg, Germany}

\author{Jan M. Pawlowski}
\affiliation{Institut
 f\"{u}r Theoretische Physik, Universit\"{a}t Heidelberg,
 Philosophenweg 16, 69120 Heidelberg, Germany}

\author{Marc Schiffer}
\affiliation{Institut
 f\"{u}r Theoretische Physik, Universit\"{a}t Heidelberg,
 Philosophenweg 16, 69120 Heidelberg, Germany}
 
 \author{Fleur Versteegen}
\affiliation{Institut
 f\"{u}r Theoretische Physik, Universit\"{a}t Heidelberg,
 Philosophenweg 16, 69120 Heidelberg, Germany}

\title{Asymptotic safety, string theory and the weak gravity conjecture}

\begin{abstract}
  We propose a scenario with string theory in the deep ultraviolet, an
  intermediate asymptotically safe scaling regime for gravity and
  matter, and the Standard Model in the infrared. This could provide a
  new perspective to tackle challenges of the two models: For
  instance, the gravitational Renormalization Group flow could connect
  a negative microscopic to a positive macroscopic cosmological
  constant, potentially rendering string theory on an anti-de Sitter
  background observationally viable. Further, the unitarity of a
  string-theoretic ultraviolet completion could be inherited by an
  asymptotically safe fixed point, despite the presence of
  higher-order interactions. We discuss necessary conditions on the
  scale of asymptotic safety and the string scale for our scenario to
  be viable. As a first test, we explore the weak-gravity conjecture
  in the context of asymptotically safe gravity.
 
\end{abstract}

\maketitle

\section{Asymptotic safety and string theory}
\label{sec:intro}

What is the fundamental nature of the building blocks of our universe?
String theory and the asymptotically safe Standard Model (ASSM) are
both possible candidates. The latter relies on scale-symmetry kicking
in at microscopic distance scales. If realized, it provides a
predictive quantum field theory of the Standard Model plus quantum
gravity, see
\cite{Shaposhnikov:2009pv,Eichhorn:2017ylw,Eichhorn:2017lry,%
  Eichhorn:2018whv,Eichhorn:2018ydy} for RG studies and
\cite{Eichhorn:2017egq} for a recent review. It is based on an
interacting fixed point of the Renormalization Group (RG),
generalizing the concept of asymptotic freedom to a setting in which
both gravity as well as Abelian gauge sectors could be included
without Landau poles. Compelling indications for asymptotic safety in
pure Euclidean gravity, proposed in \cite{Weinberg:1980gg}, have been
collected in
\cite{Reuter:1996cp,Falkenberg:1996bq,Souma:1999at,Reuter:2001ag,Lauscher:2001ya,%
  Litim:2003vp,Codello:2006in,Codello:2008vh,%
  Benedetti:2009rx,Niedermaier:2010zz,Manrique:2010am,Manrique:2011jc,%
  Dietz:2012ic,Donkin:2012ud,Codello:2013fpa,Falls:2013bv,Becker:2014qya,%
  Christiansen:2014raa,Demmel:2015oqa,Percacci:2015wwa,Gies:2016con,%
  Biemans:2016rvp,Denz:2016qks,Christiansen:2017bsy,Knorr:2017fus,%
  Knorr:2017mhu,Falls:2017lst,Falls:2018ylp,Knorr:2018fdu,%
  Pagani:2019vfm,Knorr:2019atm}, starting from the pioneering work
\cite{Reuter:1996cp}, and matter-gravity systems have been explored,
see, e.g.,
\cite{Shaposhnikov:2009pv,Eichhorn:2017ylw,Eichhorn:2017lry,%
  Narain:2009fy,Dona:2012am,Dona:2013qba,Meibohm:2015twa,Dona:2015tnf,%
  Eichhorn:2016vvy,Biemans:2017zca,Christiansen:2017cxa,Alkofer:2018fxj,%
  Eichhorn:2018akn,Eichhorn:2018nda,Eichhorn:2018whv,%
  Eichhorn:2018ydy,Wetterich2019,Bosma:2019aiu}.
For related Monte Carlo studies see e.g.\
\cite{Laiho:2016nlp,Ambjorn:2019lrm} and references therein.
Potential cosmological implications are reviewed in
\cite{Bonanno:2017pkg} and possible consequences for black-hole
physics explored, e.g., in
\cite{1999alfiomartin,Bonanno:2000ep,Falls:2010he,Litim:2013gga,%
  Torres:2017ygl,Pawlowski:2018swz,Adeifeoba:2018ydh,Platania:2019kyx,Held:2019xde},
see \cite{Percacci:2017fkn,Reuter:2019byg,Pereira:2019dbn} for recent
reviews. See \cite{Wetterich:2019qzx} for an introduction to quantum
scale symmetry and \cite{Eichhorn:2018yfc} for a review of asymptotic
safety and underlying mechanisms in various models.

The transition scale $k_{\rm tr}$ is defined as the energy scale at
which a departure from scale-symmetry sets in, such that below
$k_{\rm tr}$ the couplings deviate from their values in the
asymptotically safe scaling regime. The additional scale-symmetry in the
asymptotically safe regime can even result in enhanced predictive
power, potentially fixing the values of some of the Standard-Model
couplings at $k=k_{\rm tr}$, and thereby at all scales
\cite{Shaposhnikov:2009pv,Eichhorn:2017ylw,Eichhorn:2017lry,%
  Eichhorn:2018whv,Eichhorn:2018ydy}.  For our scenario, it is key
that the determination of infrared (IR) values of couplings following
from scale-symmetry also carries over (at least approximately) if an
asymptotically safe scaling regime is only realized over a finite, but
large enough, range of scales, with new physics kicking in at very
high energy scales.

\begin{figure}[!t]
  \includegraphics[width=\linewidth, clip=true, trim=3cm 2cm 5cm
  13cm]{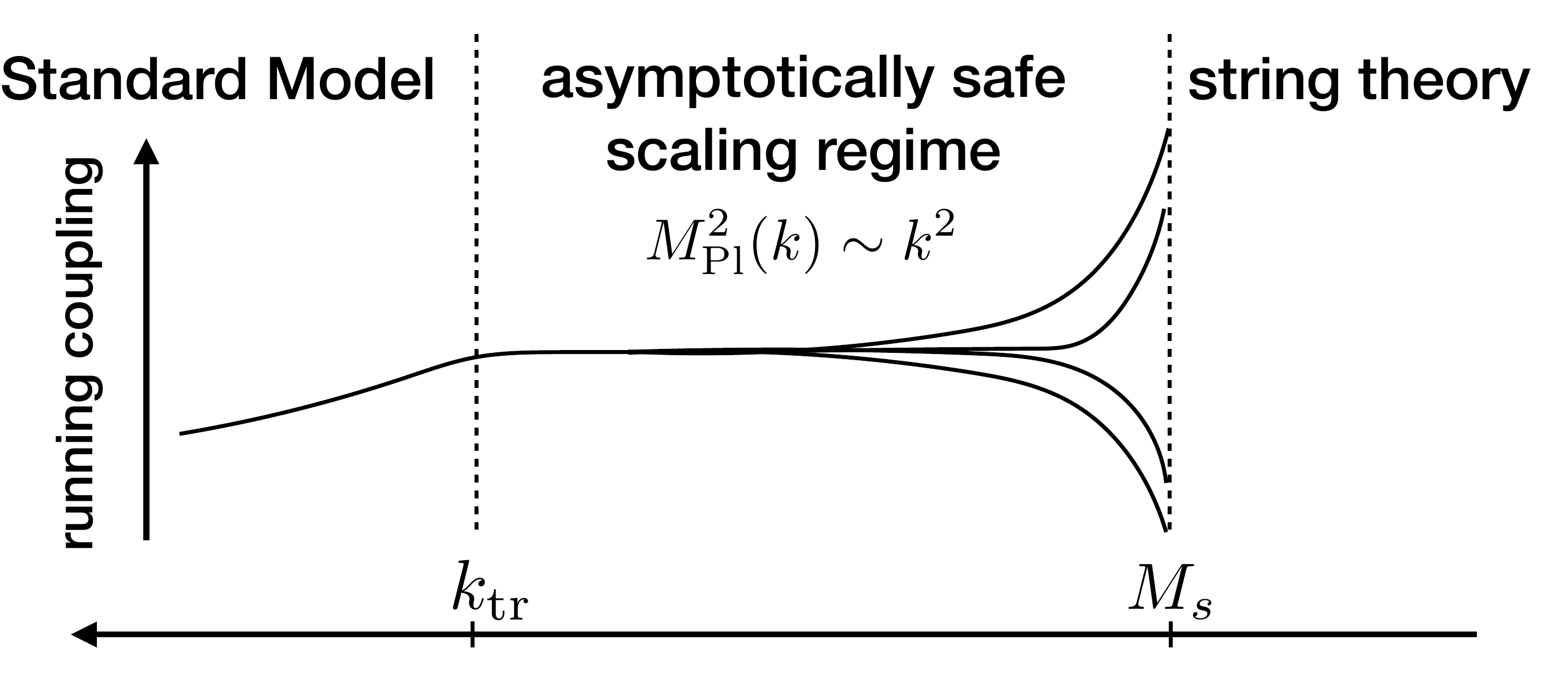}
\caption{\label{fig:ill_scales} We illustrate our scenario, indicating
  how an asymptotically safe scaling regime can generate universal
  predictions for couplings, coming from a range of values resulting
  from different choices of compactification for the string theory at
  the string scale $M_s$. The scale $k_{\rm tr}$ is the transition
  scale from an asymptotically safe scaling regime, where relevant
  operators kick in and drive the flow away from the scale-invariant
  point.}
\end{figure}

\begin{figure}[!t]
  \includegraphics[width=\linewidth, clip=true, trim=2cm 2cm 12cm
  5cm]{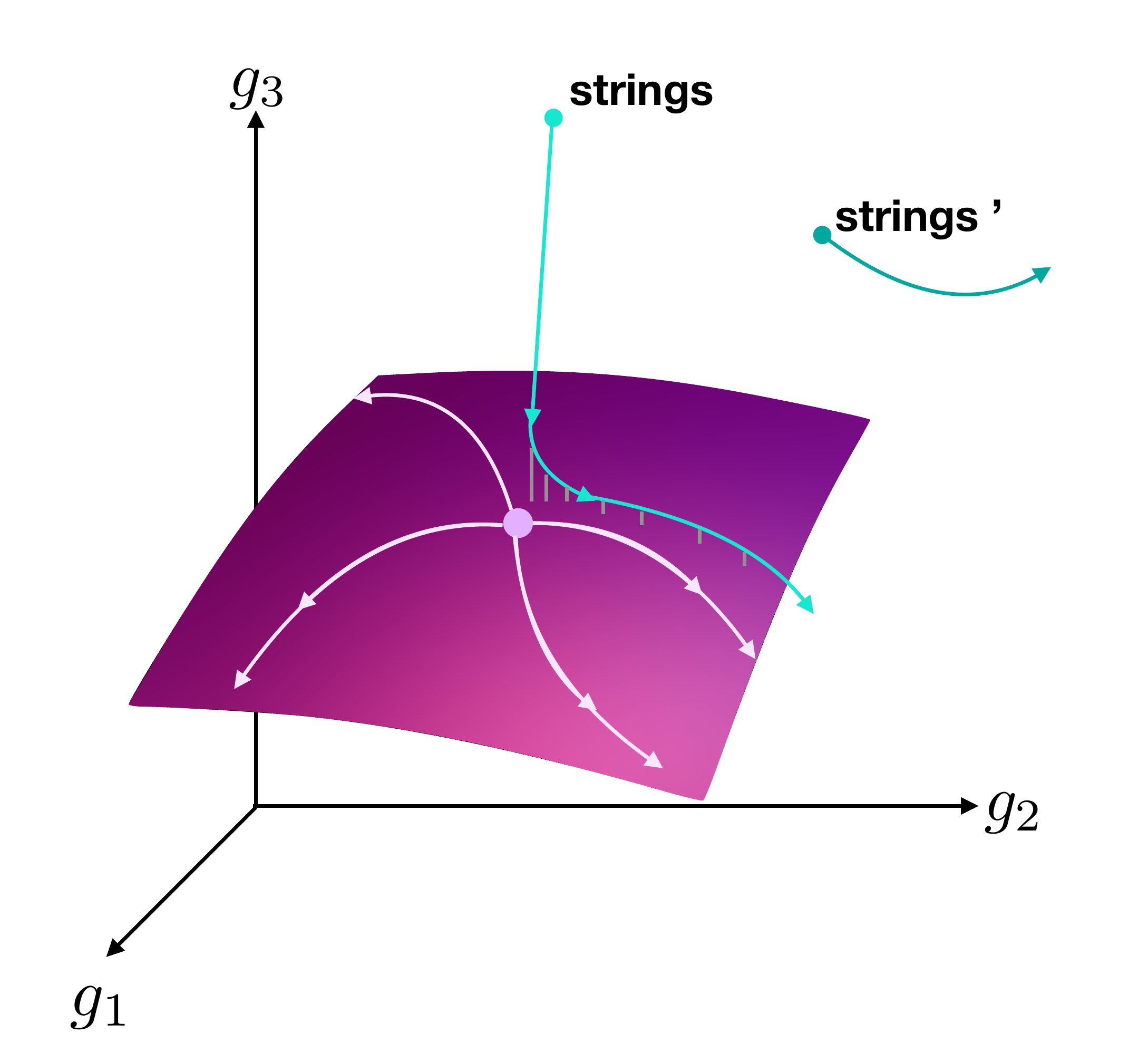}
  \caption{\label{fig:string_AS_illu} We show a sketch of a
    three-dimensional space of couplings with an asymptotically safe
    fixed point (light purple) and its UV critical surface
    (purple). Its IR critical surface is one-dimensional, and the
    starting point provided by a string model (light cyan) lies within
    it, resulting in the effective QFT description of this string
    model approaching the fixed point very closely, before the RG
    trajectory leaves the fixed-point regime close to the UV critical
    surface. For an alternative string model (string ') the starting
    point for the QFT description (darker cyan) lies off the IR
    critical surface of the
    fixed point. \\
    An earlier discussion explaining how models which are not
    fundamentally asymptotically safe can nevertheless appear
    effectively asymptotically safe can be found in
    \cite{Percacci:2010af}.}
\end{figure}

On the other hand, string theory goes beyond the local quantum field
theory framework, resulting in the requirement for extra dimensions as
well as supersymmetry, see, e.g.,
\cite{Polchinski:1998rq,Polchinski:1998rr, Becker:2007zj} for reviews.
Both the transition scale $k_\text{tr}$ in asymptotic safety and the
string scale $M_\text{s}$, are usually associated with the Planck
scale. Therefore, a relation between these two candidates for a
microscopic description of nature might not be immediately obvious,
but could actually be possible whenever these two scales are
separated, i.e., $k_\text{tr} < M_\text{s}$. Here, we set out to
investigate a possible connection. We refer to
Fig.~\ref{fig:ill_scales} for an illustration of our proposal.
Specifically, the scenario we explore assumes that string theory
provides the most fundamental description of nature.
Below the string scale $M_s$, this results in an effective
quantum-field theoretic description\footnote{For simplicity, we are
  taking the compact volume to be not that large so that the
  Kaluza-Klein (compactification) scale is close to the string
  scale.}. We assume that the values of couplings at $M_s$ lie in the
IR basin of attraction of the asymptotically safe fixed point. This
assumption results in constraints on those couplings that are
\emph{relevant} at the interacting fixed point, as those are the
IR-repulsive directions, cf.~Fig.~\ref{fig:string_AS_illu}.  Along the
IR attractive (irrelevant) directions of the fixed point, the flow is
pulled towards the fixed point. This results in an RG trajectory that
spends a large amount of RG ``time" close to the fixed point and then
emanates from its vicinity close to the UV critical
surface.  In the simplest case, the compactification scale and scale
of supersymmetry breaking are both close to the string scale $M_s$, so
that the effective field theory is four-dimensional, potentially
facilitating an asymptotically safe fixed point for all gauge
interactions, including an Abelian one \cite{Eichhorn:2019yzm}.   For the simplest scenario, we also assume that additional states from string theory (such as heavy moduli and superpartners) decouple at high energies (i.e., directly below the string scale ), so that the effective-field theory regime contains only the Standard Model and gravity. This assumption can be relaxed to accommodate further matter fields that arise from string theory, if an asymptotically safe fixed point persists under extensions by the corresponding additional fields. We will work with general numbers of matter fields (which may include light moduli, axions, etc) in the following.
In
brief, our setup explores those parts of the string landscape that
feature an emergent scale-symmetry.

The degree to which the asymptotically safe scaling regime determines
the deep-IR physics by mapping a given range of initial conditions at
the string scale to a narrow IR range of couplings,
cf.~Fig.~\ref{fig:ill_scales}, depends on the following two
properties:
  \begin{itemize}
  \item[(i)] How strongly the irrelevant couplings are attracted to
    the asymptotically safe fixed point.
  \item[(ii)] How large the separation is between the asymptotically
    safe transition scale $k_{\rm tr}$ and the string scale $M_s$.
\end{itemize}

In such a setting, the physics in the deep IR is essentially
determined by the ASSM. This might include the intriguing consequence
that the Higgs mass
\cite{Shaposhnikov:2009pv,Eichhorn:2017als,Pawlowski:2018ixd},
the top quark mass \cite{Eichhorn:2017ylw}, the bottom quark
mass \cite{Eichhorn:2018ydy} and the Abelian gauge coupling
\cite{Harst:2011zx,Eichhorn:2017lry} could emerge as predictions of
string theory.  This follows since functional RG studies indicate that the
  respective couplings come out as \emph{irrelevant} couplings with
finite asymptotically safe fixed-point values.  In turn, the relevant couplings in state-of-the-art
approximations in asymptotically safe gravity are the cosmological
constant, the Newton coupling, and a superposition of the 4-derivative
curvature couplings, see, e.g.,
\cite{Benedetti:2009rx,Falls:2013bv,Denz:2016qks}.  Thus, the
constraint of reaching this fixed point with the given relevant
couplings from string theory selects a highly predictive corner of the
string landscape.

We also point out that the gravitational RG flow can connect a
fixed-point regime at negative cosmological constant to an IR regime
with a tiny, positive value of the cosmological constant, as required
observationally. This could help to address a challenge in string
theory, where the existence of consistent de Sitter (dS) backgrounds
such as the $\bar {D}$ (anti-D-brane) uplift of KKLT
\cite{Kachru:2003aw}, see also \cite{Cicoli:2012fh,Cicoli:2015ylx,Gallego:2017dvd} for other
  potential constructions, is under debate, see, e.g.,
\cite{Moritz:2017xto,Kallosh:2018nrk,Akrami:2018ylq,%
  Cicoli:2018kdo,Kallosh:2019axr,Hamada:2019ack}. Typically, in string
theory, it is more natural to get anti-de Sitter (AdS) backgrounds,
and in contrast to the supersymmetric AdS background of KKLT (prior to
introducing $\bar{D}$'s) one can even get SUSY broken AdS backgrounds
\cite{Balasubramanian:2005zx}. In most string phenomenology
discussions based on the latter, some additional input (not
necessarily $\bar{D}$'s, see e.g., \cite{Cicoli:2012fh,Cicoli:2015ylx,Gallego:2017dvd}) is used to 'uplift' such an AdS minimum to
dS. However, this last step is somewhat less well under control
compared to the original AdS construction in
\cite{Balasubramanian:2005zx}.

The difficulty of getting stable dS vacuum configurations in string
theory (see \cite{Danielsson:2018ztv} for a recent discussion) has
even been elevated to the level of a conjecture \cite{Obied:2018sgi}
\cite{Ooguri:2018wrx}, see also \cite{Garg:2018reu}, stating that it
is not possible to get a stable dS solution in a controlled
approximation scheme within string theory. Be that as it may (and in
fact, this conjecture is indeed controversial, see, e.g.,
\cite{Hebecker:2018vxz}), it should be pointed out that while the
effective field theory arising from string theory is expected to be
defined at (or close to) the string scale, the observed positive
cosmological constant is measured in the deep IR. Hence it is
conceivable that a negative cosmological constant obtained from string
theory is consistent with a positive cosmological constant at
cosmological scales.  It should be emphasized that the cosmological
constant is an IR repulsive coupling of the asymptotically safe
scaling regime. Hence, an RG trajectory which realizes such an AdS-dS
transition is not generic but has to be set by rather specific initial
conditions of the effective field theory arising from string
theory. Nevertheless, it could connect a string theory with a negative
microscopic cosmological constant to a positive cosmological constant
in the infrared.

Establishing a relation between asymptotically safe gravity and string
theory is also interesting for the following reason: The presence of a
fixed point of the RG flow is not sufficient to guarantee a
well-defined ultraviolet completion, as the microscopic dynamics might
feature kinematical instabilities, leading to a unitarity problem.
Four-derivative gravity, which features an asymptotically free UV
completion \cite{Stelle:1976gc}, is typically considered an example of
the fact that the presence of higher-derivative terms can spoil
unitarity. Note, however, that the mere existence of kinematic
instabilities at the classical level or at a finite order of
derivatives is far from being conclusive with regard to unitarity or
its lack. The possibility of non-perturbative cures of perturbative
unitarity problems has been investigated recently in e.g., 
\cite{Holdom:2016xfn,Donoghue:2017fvm,Becker:2017tcx,Anselmi:2018ibi}.

Asymptotically safe gravity is an example of a non-perturbative setup
and higher derivative terms typically come to all orders. Thus, a
Taylor expansion of the inverse propagator up to finite order in
momenta (which generically features additional zeros) is inadequate to
answer the question if asymptotically safe gravity is unitary. For a
recent discussion of this see, e.g., \cite{deAlwis:2018azs}. Of
course, in turn, it makes a conclusive analysis even more intricate.

Within the scenario we explore here, the above intricacies are
softened: Additional poles in the gravity propagator can be present
without spoiling the consistency of the theory, as long as they lie at
or beyond the string scale. Conversely, within the present scenario
one can even use the scale of additional poles in order to estimate
the required value of the fundamental string scale.

In Sec.~\ref{sec:conditions}, we present explicit conditions on the
parameters of the fundamental string theory and the intermediate
asymptotically safe scaling regime that are necessary to realize the
required separation of scales, i.e., $k_\text{tr}<M_s$,
cf.~Fig.~\ref{fig:ill_scales}. In Sec.~\ref{sec:AdS-dS}, we comment on
the possibility of a transition from a negative cosmological constant
at $M_s$ transitioning to a viable positive value at macroscopic
scales. In Sec.~\ref{sec:WGC}, we discuss first implications of the
weak gravity conjecture in the presented scenario. Finally, we
summarize and give an outlook in Sec.~\ref{sec:conclusions}.

\section{Conditions realizing an intermediate scaling regime}
\label{sec:conditions}

Let us now analyze the conditions on the string scale $M_s$ and the
transition scale $k_{\rm tr}$ that have to hold within our proposed
scenario. These considerations can inform model-building efforts, both
on the string-theory side as well as the asymptotically safe side.  To
that end, we now discuss the flow of the gravitational
coupling. Define the dimensionless gravitational coupling at the
momentum scale $k$ as
\begin{equation}
    g(k)\equiv\frac{k^{2}}{8\pi M_{{\rm Pl}}^{2}(k)}\,.\label{eq:gk}
\end{equation}
Here $M_{{\rm Pl}}(k$) is the running Planck scale - the physical
gravitational coupling giving the initial condition in the deep IR is
$M_{{\rm Pl}}^{2}(k=0)\equiv 1/(8\pi\, G_{\rm{Newton}})$.  To lowest order
in the truncation of the infinite series for the beta function of the
gravitational coupling we have in a semi-perturbative approximation \footnote{In this approximation higher dependencies of the coupling $g$ that come from the fully non-perturbative propagator are neglected. Applying this approximation to marginal couplings allows us to recover the universal 1-loop coefficients, see, e.g., \cite{Gies:2006wv}.}
\begin{equation}
  \beta_{{\rm gravity}}=\frac{dg}{dt}=2g-2 \frac{g^{2}}{g_{*}}\,.\label{eq:betag}
\end{equation}
Here $t=\ln k$ and $g_{*}$ is the fixed point value of $g$. The
fixed-point coupling $g_*$ needs to be positive in order to have a
physically meaningful asymptotically safe theory. For a UV fixed point
with $g_*< 0$ the fixed point at $g_{*}=0$ shields the UV fixed point
from a low-energy regime with attractive gravity, as realized in our
universe.  In pure gravity the UV fixed point has been found at
$g_*>0$,
\cite{Reuter:1996cp,Falkenberg:1996bq,Souma:1999at,Reuter:2001ag,Lauscher:2001ya,%
  Litim:2003vp,Codello:2006in,Codello:2008vh,%
  Benedetti:2009rx,Niedermaier:2010zz,Manrique:2010am,Manrique:2011jc,%
  Dietz:2012ic,Donkin:2012ud,Codello:2013fpa,Falls:2013bv,Becker:2014qya,%
  Christiansen:2014raa,Demmel:2015oqa,Percacci:2015wwa,Gies:2016con,%
  Biemans:2016rvp,Denz:2016qks,Christiansen:2017bsy,Knorr:2017fus,%
  Knorr:2017mhu,Falls:2017lst,Falls:2018ylp,Knorr:2018fdu,%
  Pagani:2019vfm,Knorr:2019atm}. This is a consequence of
gravitational fluctuations having an antiscreening effect on the
Newton coupling, thereby generating an asymptotically safe fixed-point
regime. Of course, matter fluctuations also drive the value of $g_*$,
towards either larger or smaller values, as has been explored in
\cite{Dona:2013qba,Meibohm:2015twa,Dona:2015tnf,%
  Biemans:2017zca,Christiansen:2017cxa,Alkofer:2018fxj,%
  Eichhorn:2018ydy,Eichhorn:2018nda,Wetterich2019,Bosma:2019aiu}. In a
first, rough, approximation we may write this dependence of $g_*$ as
\begin{equation}
     g_{*}(N_\textrm{eff})\approx \frac{12\pi}{N_{\rm eff}}\,.\label{eq:g*}
\end{equation}
Roughly speaking, $N_\textrm{eff}$ comprises a weighted sum of the
number of spin $s$ fields with $s=0,1/2,1,3/2$ and contains the effect
of metric fluctuations, $s=2$. The higher spin modes (see
\cite{Dona:2014pla}) are required for supersymmetric extensions of the
Standard Model.  The detailed fixed-point properties of fully coupled
gravity-matter system -which contain higher-order as well as
non-minimal interactions- is subject of current research.

We proceed with the discussion of the consequences of this setup. We
first focus on the case $g_*(N_\textrm{eff})> 0$, that is $N_{\rm
  eff}>0$, and comment on the second case below.  Integrating the flow
equation \eqref{eq:betag} and re-expressing in terms of the running
Planck scale, cf. equation~\eqref{eq:gk}, we have
\begin{equation}
    M_{{\rm Pl}}^{2}(k)=M_{{\rm Pl}}^2(0)+\frac{1}{8\pi g_{*}}k^{2}\,.\label{eq:Plflow}
\end{equation}
where $M_{\rm Pl}^2(0)$ is the low-energy Planck mass, i.e., we have
set the low-energy reference scale $k_0=0$.  For
$k^2\ll 8\pi M_{\rm Pl}^2(0)\, g_{*}$, the dimensionful Planck mass is
essentially constant, $M^2_{\rm Pl}(k) \approx M_{\rm Pl}^2(0)$, as
expected in the classical-gravity regime.  In contrast, for
$k^2> 8\pi M^2_{\rm Pl}(0)\, g_{*}$, we are in the asymptotically safe
scaling regime, where the Planck mass exhibits scaling,
$M_{\rm Pl}(k^2)\sim k^2$.

At the transition scale $k=k_{\rm tr}$, the scale-dependence vanishes,
such that the following estimate for the transition scale holds
\begin{equation}
k_{\rm tr}^2 =8\pi\, M_{\rm
    Pl}^2(0)\,g_{\ast}\,.\label{eq:ktrans} 
\end{equation}
If the fixed-point value is sufficiently low, fixed-point scaling can
even set in well below the Planck scale. For $g_{*}\sim O(1)$ the
quantum correction to the running (squared) Planck scale is a small
($\sim4\%)$ effect even at $k=M_{{\rm Pl}}(0)$. But, if
$N_{\rm eff}\gg1$, so that $g_{*}\ll1$, the quantum corrections can be
significant.  Such a change of the fixed-point value of the Newton
coupling could follow from the impact of quantum fluctuations of
matter, see e.g., \cite{Dona:2013qba,
  Meibohm:2015twa,Biemans:2017zca}. Whether this is indeed realized
with a suitable number of matter fields is beyond the scope of the
present work. 

In view of the flow equation \eqref{eq:Plflow}, one needs to
reconsider the relation between the matching scale $\bar{k}$, at which
QFT should be replaced by string theory, and the low-energy Planck
scale $M_{{\rm Pl}}(0)$.  If an asymptotically safe scaling regime is
realized, the matching relations should actually use the running
Planck scale, cf.~Eq.~\eqref{eq:gk}, at the matching scale $\bar{k}$,
which differs from the low-energy Planck scale
\begin{equation}
  M_{{\rm Pl}}^{2}(\bar{k})=\frac{\bar{k}^{2}}{8\pi g(\bar{k})}=
  \frac{M_{s}^{2}{\cal V}}{\sqrt{g_{s}}}\,,\label{eq:MP}
\end{equation}
where ${\cal V}$ is the volume of the compact space in string units,
and $g_s$ is the string coupling. This is because the relation between
the 4D Planck scale and the string scale is expected to be valid at
the cutoff scale which we denote by $\bar{k}$. This relation can be
read off from the low-energy effective action.

The matching scale $\bar{k}$ should be somewhat less than the
Kaluza-Klein(KK) scale, which is related to the string scale by
$M_{{\rm KK}}^{2}=M_{s}^{2}/{\cal V}^{1/3}$.  Using Eq.~\eqref{eq:MP}
to solve for the string scale we arrive at
\begin{equation}
  M_{{\rm KK}}^{2}=\frac{\sqrt{g_{s}}}{{\cal V}^{4/3}}
  \frac{\bar{k}^{2}}{8\pi g(\bar{k})}\gtrsim\bar{k}^{2}\,.
\end{equation}
This gives the bound on the compact space volume
\begin{equation}
  \frac{{\cal V}^{4/3}}{\sqrt{g_{s}}}\lesssim\frac{1}{8\pi g(\bar{k})}<
  \frac{1}{8\pi g_{*}}\frac{M_{{\rm Pl}}^{2}(\bar{k})}{M_{{\rm Pl}}^{2}(0)}\,.\label{ineq:1}
\end{equation}
 The second inequality comes from the requirement that there is
  a scaling regime, i.e., that 
  \be 
  \bar{k}^2>k_{\rm tr}^2.  
  \ee  
  In
full theory space (i.e., the space of gravitational couplings), the
existence of the scaling regime depends on the values of the
gravitational couplings at $\bar{k}$ in relation to the location of
the asymptotically safe fixed point. For simplicity we now strengthen
the inequality \eqref{ineq:1} by neglecting the flow of all other
couplings.

Now from Eq.~\eqref{eq:Plflow} we have 
\begin{equation}
  \frac{M_{{\rm Pl}}^{2}(\bar{k})}{M_{{\rm Pl}}^{2}(0)}=\left(1-
    \frac{g(\bar{k})}{g_{*}}\right)^{-1}\,.
\end{equation}
This shows that for a long scaling regime, $g(\bar{k})$ should be
sufficiently close to $g_{\ast}$, just as one would expect.

We may rewrite the second inequality in \eqref{ineq:1} further as
\begin{equation}
\frac{1}{g(\bar{k})}<\frac{1}{g_{*}-g(\bar{k})}\,.
\end{equation}
Further using that $g_{*}>0$, this implies that
\begin{equation} 
1<\frac{g_{*}}{g(\bar{k})}<2\,,
\label{eq:g-inequality}
\end{equation}
where the first inequality comes from the requirement of positivity of
$g(\bar{k})$.  The first inequality also ensures that the
  potential scaling regime connects the string theory to the Gaussian
  fixed point, i.e., to a viable IR limit.  Thus, the bound on the
volume in \eqref{ineq:1} can be expressed in terms of the fixed- point
value of the dimensionless gravitational coupling as
\begin{equation}
  \frac{{\cal V}^{4/3}}{\sqrt{g_{s}}}\lesssim\frac{1}{8\pi g(\bar{k})}<
  \frac{2}{8\pi g_{*}}\,.\label{eq:A-N}
\end{equation}
The two inequalities \eqref{eq:A-N} and
    \eqref{eq:g-inequality} together ensure that there is a scaling
  regime, i.e., that $\bar{k}^2>k_{\rm tr}^2$, and that it
    connects to a viable IR limit.
Accordingly these inequalities can be satisfied by either
\begin{enumerate}
	\item a small asymptotically safe fixed-point value for $g_\ast$,
	\item or a large string coupling $g_s$.
\end{enumerate}
Note that a third possibility i.e., ${\cal V} < 1 $ is not realizable
because of T-duality considerations.  This essentially means that the string scale is a lower limit for length scales - smaller scales have to be analysed in terms of the T-dual theory. For instance a type IIB compactification with a Calabi-Yau space with some Euler character $\chi$ at below the string scale is actually a type IIA theory with Euler character $-\chi$. Thus one simply has to replace one string compactification model by another. (See for instance the discussion in \cite{Polchinski:1998rq,Polchinski:1998rr} chapters 8 and 13.)
Given either of
these conditions the proposed scenario summarized in
Fig.~\ref{fig:ill_scales} might be realized. Let us now comment on
them further.

The first option for satisfying Eq.~\ref{eq:A-N} is a fixed-point
value of the Newton coupling which is sufficiently small. In such a
setting $\bar{k}$ might even be as low as the infrared Planck scale,
while $k_{\rm tr}^2<M_{\rm Pl}^2(0)$ would need to hold.  This would
imply a weakly coupled asymptotically safe regime with a very small
fixed-point value. It is intriguing that hints for a rather
weakly-coupled (in the sense of near-Gaussian scaling behavior)
asymptotically safe regime have been found in pure gravity
\cite{Falls:2013bv,Falls:2014tra,Falls:2018ylp}, and in particular
with matter
\cite{Eichhorn:2018akn,Eichhorn:2018ydy,Eichhorn:2018nda}. The latter
also might allow a near-perturbative UV completion for the Standard
Model
\cite{Eichhorn:2017ylw,Eichhorn:2017lry,Eichhorn:2018whv,Eichhorn:2019yzm}.
Such a scenario might be achievable under the impact of an appropriate
number and type of matter degrees of freedom
\cite{Dona:2013qba,Meibohm:2015twa,Dona:2015tnf,%
  Biemans:2017zca,Christiansen:2017cxa,Alkofer:2018fxj,%
  Eichhorn:2018ydy,Eichhorn:2018nda,Wetterich2019}.
  
For the second option, the string theory would have to be strongly
coupled, i.e., $g_s$ could be sufficiently large. While this is not
necessarily a regime that is computationally easy to access on the
string side, it is nevertheless intriguing to observe that the
strongly-coupled string regime could be related to a weakly-coupled
asymptotically safe regime in our setting. However often a strongly
coupled string theory is S-dual to another string theory in the weak
coupling regime - for instance, type I string theory in strong/weak
coupling is S-dual to heterotic SO(32) string theory, while type IIB
string theory is self-dual under S-duality (in effect SL(2,Z))
transformations. Hence if one finds that a given asymptotically
safe field theory is related to a strongly coupled regime of the
corresponding string theory, the latter should be replaced by its
S-dual weakly coupled partner and the corresponding field theory
looked at for its asymptotically safe properties \footnote{We wish to
  thank Arthur Hebecker for drawing our attention to this issue.} . In
the case of type IIB not only the string coupling, but also 
the fluxes in the compactification manifold change, thus 
changing the phenomenology.

In the other case, $N_{\rm eff}<0$, there is no
asymptotic safety, since $g_{*}<0$, and the cutoff scale  
following from the running of $g$ in Eq.~\eqref{eq:betag}
is
\begin{align}
\bar{k}^{2}
\simeq&\frac{96\pi^{2}M_{\rm{Pl}}^{2}(0)}{|N_{\rm eff}|}\ll M_{\rm{Pl}}^2(0)\,,
\end{align}
where the inequality holds for large $|N_{\rm eff}|$. This scale is
basically the so-called species scale
$k_{{\rm species}}\sim M_{\rm{Pl}}(0)/\sqrt{N}$ (see for instance
\cite{Palti:2019pca}, especially the argument around eqn.~5.16).

The sign of $N_{\rm eff}$ is crucial. If it is positive, we could have
asymptotic safety and the above arguments for a potential
compatibility with string theory would be valid. In this case the UV
Planck mass may be much larger than the IR Planck mass and gravity is
weakly coupled in the UV. On the other hand for $N_{\rm eff}<0$ the UV
Planck mass is much smaller, so gravity becomes strongly coupled in
the UV. Of course all of the above arguments are strictly valid only
in the leading order truncation of the RG equations.

The above discussion would mean that the existence of an
asymptotically safe fixed point would (approximately) determine the
infinite set of irrelevant couplings at the string/cutoff/KK scale. In
string theory terms it would mean that the bottom up physics is fixing
the particular compactification, the choice of the Calabi-Yau
manifold, set of fluxes etc., i.e., a particular string theory vacuum
from the landscape.

\begin{figure}
\includegraphics[width=\linewidth]{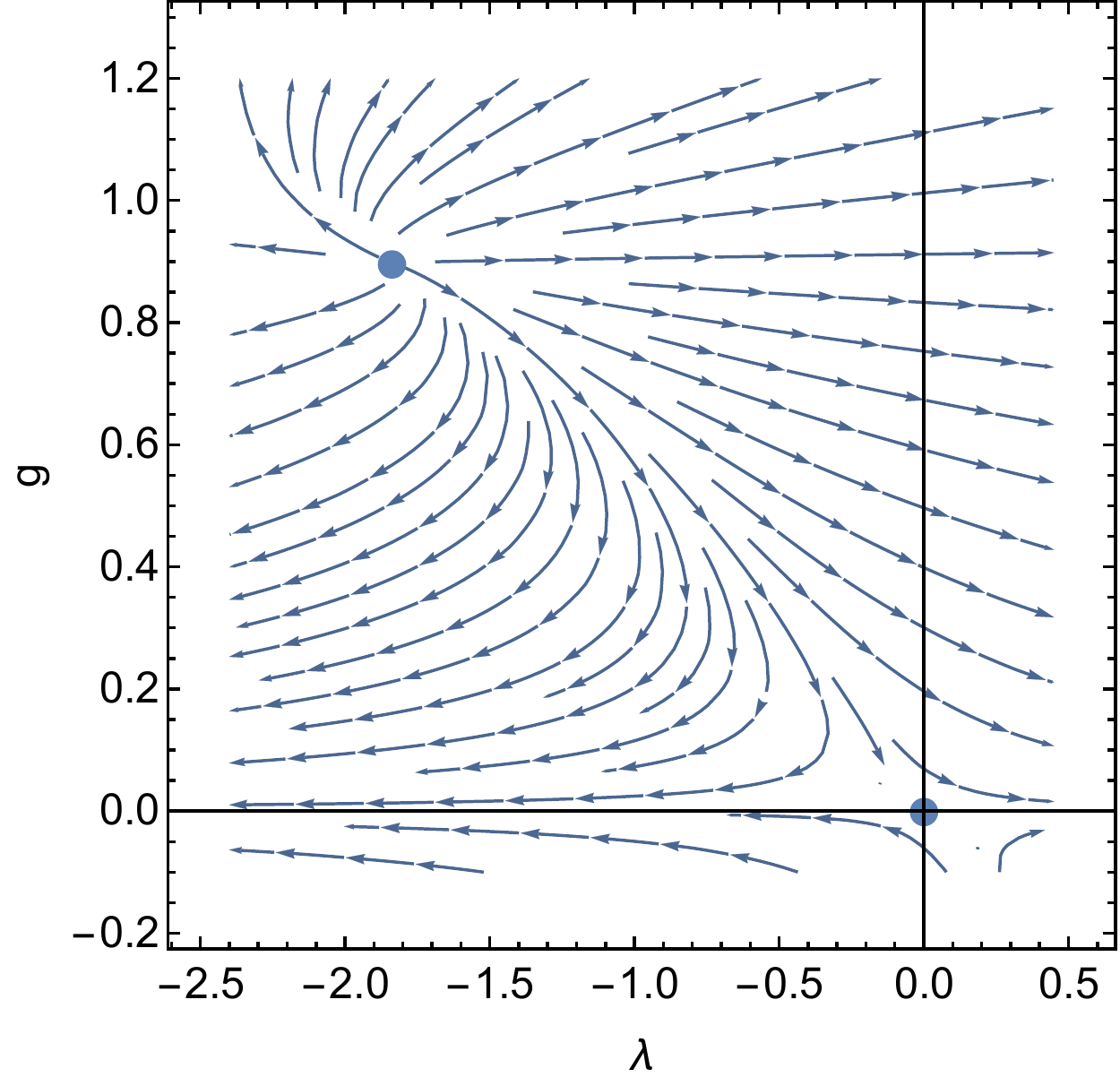}
\caption{\label{fig:flow} We show the RG flow towards the IR in the
  $(g-\lambda)$ plane in the approximation of \cite{Dona:2013qba},
  which exhibits RG trajectories crossing from the fixed point at
  negative cosmological constant to a tiny positive cosmological
  constant in the infrared for $N_{\rm eff} = 42$ and
  $N_{\rm eff}' = 66$, based on Eq.~\eqref{eq:betag} and
  Eq.~\eqref{eq:betalambda}. The chosen values for $N_{\rm eff}$ and
  	$N_{\rm eff}'$ correspond to the matter content of the Standard Model in the approximation of \cite{Dona:2013qba}.}
\end{figure}

\section{From microscopic anti de Sitter to macroscopic de Sitter}
\label{sec:AdS-dS}

\begin{figure}
\includegraphics[width=\linewidth,clip=true, trim=3cm 2cm 3cm 1cm]{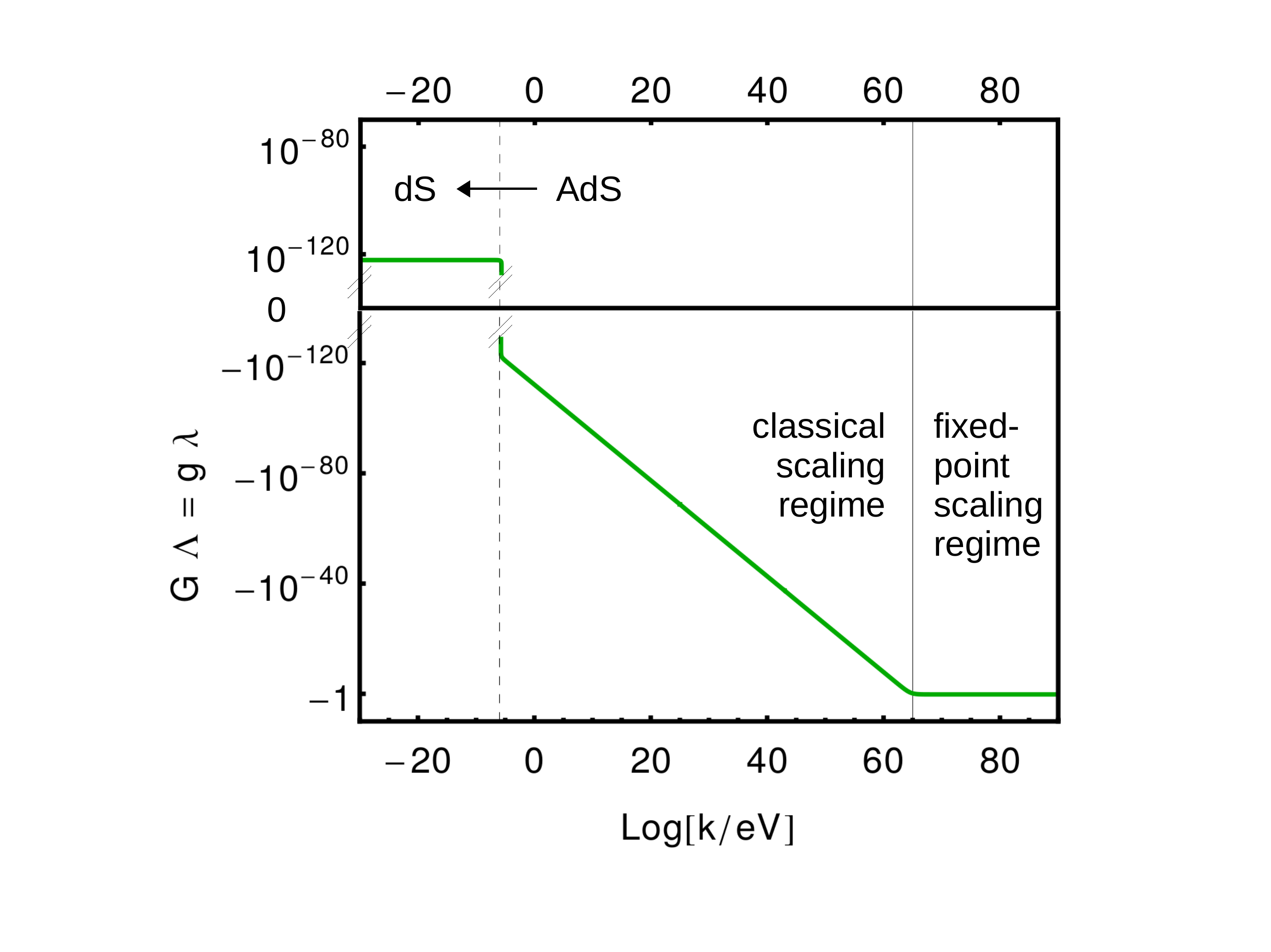}
\caption{\label{fig:AdSdS}  We show the product $G\cdot \Lambda = g\cdot \lambda$ of the dimensionfull Newton coupling $G$ and dimensionful cosmological constant $\Lambda$ along an RG trajectory that solves Eq.~\eqref{eq:betag} and \eqref{eq:betalambda} for $N_{\rm eff} = 42$ and $N_{\rm eff}' = 66$. Along the trajectory, the dimensionful cosmological constant, measured in units of the Planck mass, exhibits the asymptotically safe scaling regime in the UV, and a classical scaling regime in the IR, where it transitions from negative values (anti de Sitter) to positive values (de Sitter).}
\end{figure}

Another important property of our scenario is the dynamical change of
the cosmological constant in the UV regime.  This dynamics can turn a
negative cosmological constant at microscopic (UV) scales to a
positive one at large (IR). This happens as the cosmological constant
is not protected by symmetries in the presence of gravitational
fluctuations (i.e., at $g\neq 0$). We exemplify this in the simple
approximation of matter-gravity systems in \cite{Dona:2013qba}, where
the RG flow of the dimensionless cosmological constant
$\lambda = \Lambda/k^2$ is given by
\begin{eqnarray}
  \beta_{\lambda} &=& -2 \lambda + g\, \frac{\lambda}{6\pi}
                      \left(-N_{\rm eff} +30 \right) - \frac{g}{4\pi}N_{\rm eff}' .
                      \label{eq:betalambda}
\end{eqnarray}
Here, $N_{\rm eff}'$ and $N_{\rm eff}$ depend on the number of matter
fields. The last term in Eq.~\eqref{eq:betalambda} drives the RG flow
of $\lambda$ across $\lambda=0$ to positive values for
$N_{\rm eff}'>0$. The determination of $N_{\rm eff}'$ and
$N_{\rm eff}$ is subject to systematic uncertainties due to the choice
of truncation, see, e.g., \cite{Dona:2013qba,
  Meibohm:2015twa,Biemans:2017zca,Wetterich2019}. Working in the
approximation of \cite{Dona:2013qba}, we show the RG flow in the
$(g-\lambda)$ plane with the desired characteristics in
Fig.~\ref{fig:flow}. 
 As one can see from the flow, multiple trajectories connect the fixed-point value at negative $\lambda_{\ast}$ to a positive IR-value of the cosmological constant. Since the cosmological constant is associated with a relevant direction of the fixed point, its IR value is a free parameter, allowing us to connect a negative fixed-point value with the observed value. For an example of such a concrete  RG trajectory that is obtained as a solution to the system Eq.~\eqref{eq:betag} and \eqref{eq:betalambda}, see Fig.~\ref{fig:AdSdS}.

\section{Weak gravity conjecture}
\label{sec:WGC}

For the above scenario to be realized, requirements from string theory
for a consistent low-energy description should be satisfied by
asymptotic safety.
A prominent example is given by the weak gravity conjecture (WGC)
\cite{ArkaniHamed:2006dz}, see also, e.g.,
\cite{Banks:2006mm,Cheung:2014vva,Heidenreich:2015nta,Saraswat:2016eaz}
 and, e.g., \cite{Crisford:2017gsb,Klaewer:2018yxi} for
  applications, which states that in a theory with a $\UniG(1)$ gauge
symmetry and corresponding coupling $\eUone$, there should be a
charged particle with charge $q$ (we replace
$\eUone q \rightarrow \eUone$) and mass $\MWGC$, such that
\begin{equation}
  \eUone \MPl \geq \MWGC\, .\label{eq:WGC} 
\end{equation}
Here, $M_{\rm Pl} = M_{\rm Pl}(k)$ is the energy dependent Planck mass. In
particular, it should not be confused with the low-energy value of the
Planck mass $M_\text{Pl}(0)$ introduced earlier.

As a minimal requirement for whether asymptotically safe models could
lie in the string-theoretic landscape, we will investigate whether the
weak gravity conjecture holds in the ASSM. Within string theory,
proofs of the conjecture based on various assumptions can be found,
e.g., in \cite{Cheung:2018cwt,Hamada:2018dde,Urbano:2018kax}.

A second motivation to study the WGC in the context of asymptotic
safety is independent of string theory. Given the remoteness of the
Planck scale from experimentally directly accessible scales, direct
observational tests of quantum gravitational physics are
challenging. Accordingly, deriving restrictions on particle physics
that come from a consistent embedding into a more fundamental theory
including quantum gravity can serve as an observational guide towards
quantum gravity, highlighting the importance of understanding
  the interplay of quantum gravity with matter, as also emphasized,
  e.g., in \cite{Nicolai:2013sz}.  In this spirit, asymptotic safety
has been investigated in
\cite{Eichhorn:2011pc,Dona:2013qba,Meibohm:2015twa,Eichhorn:2016vvy,%
  Eichhorn:2017eht,Christiansen:2017cxa}. In string theory, this is
the program of delineating the landscape with respect to the swampland
\cite{Vafa:2005ui}, see \cite{Palti:2019pca} for a recent review.  As
there are more generic arguments concerning global and gauged
symmetries in black-hole spacetimes \cite{Susskind:1995da,
  Kallosh:1995hi, ArkaniHamed:2006dz, Banks:2006mm}
\cite{Harlow:2018tng}, the weak-gravity conjecture might be
expected to hold beyond string theory.  It is therefore of independent
interest to determine whether asymptotically safe gravity-matter
models obey the weak-gravity conjecture, irrespective of a possible
embedding in string theory.

To be more specific, some comments about the inequality~\eqref{eq:WGC}
are in order.  It is introduced based on actions that describe the
physics of processes at the corresponding scale.  A basic lesson from
quantum field theory is that all couplings depend on the energy scale
used to probe any physical process.  Therefore, the couplings
appearing in the inequality should be interpreted as running
couplings, as was already pointed out in the original paper
\cite{ArkaniHamed:2006dz}.  In particular, the Planck mass, which
describes the strength of gravitational interactions and the mass
$\MWGC$ depend on the energy, too. To describe this properly, we will
focus on the ratio of $\MWGC$ and Planck mass, writing the above
inequality ~\eqref{eq:WGC} as
\begin{equation}\label{eq:weakFP}
 \eUone(k)\geq \frac{\mWGC(k)}{m_{\rm Pl}(k)} \, ,
\end{equation}
where $k$ is the energy scale of the relevant physics. $\mWGC$ and
$m_{\rm Pl}$ are the dimensionless counterparts of the two
mass-scales. In particular, such dimensionless ratios of masses run,
i.e., depend on the energy scale.  In the scale-invariant,
asymptotically safe fixed-point regime, all dimensionless counterparts
of couplings are constant. This implies that in this regime
\begin{equation} 
\label{eq: WGCAS}
  e(k) = e_{\ast}\,, \quad \frac{m_{\rm Pl}(k)}{\mWGC (k)}=
  \frac{m_{\rm Pl, \, \ast}}{\mWGCast} \,.  
 \end{equation} 
 Herein, $e_{\ast}$, $m_{\rm Pl, \,\ast}$ and $\mWGCast$ are the
 fixed-point values of these couplings.  Accordingly, the fixed-point
 properties of asymptotically safe quantum gravity determine whether
 the weak-gravity conjecture holds. In the following we will only investigate a necessary condition for this, namely that  \eqref{eq:weakFP} is satisfied at the fixed point, and will not analyze whether further constraints arise along the full RG flow.

 Asymptotically safe quantum gravity is compatible with two distinct
 fixed-point structures in the matter sector, as discussed in
 \cite{Eichhorn:2017eht}. The interacting nature of gravity at an
 asymptotically safe fixed point always percolates into the matter
 sector, such that it is not possible to set all matter interactions
 to zero, as pointed out in
 \cite{Eichhorn:2011pc,Eichhorn:2012va}. Yet, marginal interactions,
 such as those in the Standard Model, as well as masses, can either be
 finite or vanishing, depending on the respective choice of one of two
 possible fixed-point structures.

 A first option is a maximally symmetric fixed point, at which only
 higher-order interactions, not relevant for our considerations, are
 present
 \cite{Eichhorn:2011pc,Eichhorn:2012va,Christiansen:2017gtg}. At this
 fixed point, all minimal gauge-interactions and
 scalar-potential-terms vanish, i.e., $\mWGCast=0$ and
 $e_{\ast}=0$. Accordingly, the scenario summarized in the inequality
 \eqref{eq:weakFP} does not apply and one would have to derive similar
 constraints for higher-order couplings.

 A second fixed point, at which $\mWGCast\neq0$ and $e_{\ast}=0$
 violates the WGC. Conversely, a fixed point at which $e_{\ast}\neq0$
 but $\mWGCast=0$ would trivially satisfy the WGC, but such
 fixed-points cannot exist, since for a charged scalar, a finite
 fixed-point value of the mass is necessarily induced by non-vanishing
 gauge interactions.
  
 The final option is a fixed point at which a finite value for the
 gauge coupling
 \cite{Harst:2011zx,Christiansen:2017gtg,Eichhorn:2017lry,Eichhorn:2018whv}
 as well as for the mass \cite{EHW2019} is realized.  A finite
 fixed-point value $e_{\ast}>0$ could be realized in asymptotically
 safe gravity-matter models. In the approximations of the dynamics in
 \cite{Daum:2009dn,Harst:2011zx,Folkerts:2011jz,Christiansen:2017gtg,%
   Eichhorn:2017lry,Christiansen:2017cxa,Eichhorn:2018whv}, it arises
 from a balance of antiscreening quantum gravity fluctuations with
 screening quantum fluctuations of charged matter, encoded in the beta
 function as follows
\begin{equation} 
\beta_{e}=-f_g\, e +\beta^{(1)}\,e^3+\mathcal{O}(e^5)\,, 
\end{equation} 
where the second term is the standard one-loop term from charged
matter. The first term arises from quantum-gravity fluctuations, and
$f_g$ depends on the gravitational couplings. Most importantly, it is
proportional to the Newton coupling, i.e., to $m_{\rm
  Pl}^{-1/2}$. Further, it depends on additional gravitational
couplings, such as the cosmological constant.  In a perturbative
setting, a similar contribution has been discussed in
\cite{Robinson:2005fj,Toms:2008dq,Toms:2010vy,Anber:2010uj,Ellis:2010rw}.
In the asymptotically safe fixed-point regime,
$m_{\rm Pl}=m_{\rm Pl, \, \ast}$, such that $f_{g}= \rm
const$. Functional RG studies yield $f_g\geq 0$
\cite{Daum:2009dn,Harst:2011zx,Folkerts:2011jz,Christiansen:2017gtg,%
  Eichhorn:2017lry,Christiansen:2017cxa,Eichhorn:2018whv}.  Hence, a
fixed point for the gauge coupling in the one-loop approximation
arises at
\begin{equation} 
    e_{\ast} = \sqrt{\frac{f_g}{\beta^{(1)}}}\,.
\end{equation}
We now distinguish between fermionic and bosonic fields as candidates
for the light, charged particle in the WGC. In the Standard Model,
fermions are protected from acquiring a mass at high energies by
chiral symmetry, even in the presence of quantum-gravity fluctuations
\cite{Eichhorn:2011pc,
  Meibohm:2016mkp,Eichhorn:2016vvy,Eichhorn:2017eht,Gies:2018jnv}. An
explicit breaking of chiral symmetry through finite fixed-point values
for Yukawa couplings is possible
\cite{Eichhorn:2016esv,Eichhorn:2017eht,
  Eichhorn:2017ylw,Eichhorn:2018whv} in conjunction with a finite
vacuum-expectation value for a scalar, leading to finite fermion
masses.  Here, we assume that no spontaneous symmetry breaking occurs
beyond the Planck scale, or to be more precise beyond
$eM_\text{Pl}$. Therefore fermions remain massless in the UV
fixed-point regime. Thus, as $e_{\ast}>0$, the weak-gravity conjecture
is trivially satisfied in this case. We conclude that asymptotically
safe models in which a light charged fermion exists, which acquires
its mass through spontaneous symmetry breaking below the Planck scale,
appear to be compatible with the weak-gravity conjecture. Accordingly,
such models could lie in the landscape of string theory.

In the following, we focus on a charged scalar field as the lightest
charged particle. As a consequence of finite fixed-point values for
the Planck mass and the gauge coupling, $\mWGCast$ must be finite, as
well. Specifically, the beta function for the mass is given by
\begin{eqnarray}
\label{eq:mass-beta-schematic}
\beta_{\mWGC^2} &=& k\,\partial_k\,\mWGC^2 \nonumber\\
                &=& -2\,\mWGC^2 +f_m\,\mWGC^2 - \frac{3}{32\pi^2}e^2 +\dots
\end{eqnarray}
It includes a canonical term $-2\,\mWGC^2$, a contribution from
gauge-field fluctuations $\sim e^2$ and a gravitational contribution
$\sim f_m$. Just as in the case of the gauge coupling, $f_m$ depends
on the gravitational couplings including the Newton coupling but also,
e.g., the cosmological constant, see, e.g.,
\cite{Narain:2009fy,Zanusso:2009bs,Vacca:2010mj,Hamada:2017rvn,Eichhorn:2017als}
for the explicit form.  For simplicity, we have omitted additional
contributions due to scalar self-interactions here.  At the
asymptotically safe fixed point, $m_{\rm Pl} = m_{\rm Pl, \, \ast}$
and $e = e_{\ast}$.  As a consequence of $e_{\ast}\neq 0$, we cannot
set $\mWGCast=0$. Instead, a finite fixed-point value for the mass is
generated, see also \cite{EHW2019},
\begin{align}\label{eq: mWGC}
  \mWGCast^2 = \frac{-3\,e_{\ast}^2}{32\pi^2\left(2-f_m\right)}\,.
\end{align}
This expression requires some explanations. Depending on $f_m$,
$\mWGCast^2$ can have either sign. A negative sign indicates a phase
of spontaneously broken symmetry.  In the following, we focus on the
simpler case $f_m>2$. The beta function
Eq.~\eqref{eq:mass-beta-schematic} already shows that the
quantum-gravity contribution acts like an effective change of
dimensionality for the mass parameter. It is positive
\cite{Narain:2009fy,Zanusso:2009bs,Vacca:2010mj,%
  Hamada:2017rvn,Eichhorn:2017als}, and can even become larger than
2. In this case, quantum-gravity fluctuations render the Higgs
mass-parameter irrelevant. This could provide a solution to the
gauge-hierarchy problem, as proposed in \cite{Wetterich:2016uxm}:
Starting from an arbitrary value of the Higgs mass at the scale
$\Lambda_{\rm string}$, quantum fluctuations of the metric drive the
mass towards zero for a sufficiently large separation between
$\Lambda_{\rm string}$ and $k_{\rm tr}$, such that it becomes
naturally tiny at the Planck scale. This solution to the
gauge-hierarchy problem also becomes available for those string models
for which asymptotic safety is the effective low-energy description.
We highlight that the present solution only requires new physics at
the Planck scale. This is unlike most solutions to the hierarchy
problem, which require new physics close to the electroweak scale. The
key point about the resurgence mechanism is that the new physics -- in
this case quantum gravity -- provides a very particular microscopic
value of the Higgs mass parameter at the Planck scale, such that it is
automatically much smaller than the Planck scale, even though it
depends on the cutoff scale quadratically below the Planck scale.  For
this scenario, $f_m>2$ must hold such that the fixed-point value for
the mass is positive. Accordingly, the weak-gravity conjecture becomes
a nontrivial constraint on the asymptotically safe theory, as we will
show now.

Inserting the fixed-point value \eqref{eq: mWGC} for the mass
$m_{\rm WGC}$, the fixed-point value for the charge actually drops out
of the inequality \eqref{eq:weakFP}, to wit 
\begin{equation}
  g_{\ast} \leq
  \frac{4\pi}{3}\left(f_m-2\right)\,.
  \label{ineq:finalWGC}
\end{equation}
Herein, we have used the relation between Newton coupling and Planck
mass, $g= 1/(8\pi m^2_{\rm Pl})$.  The inequality
\eqref{ineq:finalWGC} actually constitutes a nontrivial constraint on
the microscopic gravitational parameter space, since $f_m$ depends on
$g$ as well as additional gravitational couplings.  In the simplest
approximation, this becomes a restriction on the microscopic value of
the cosmological constant. Given this restriction on parameter space,
one can check whether an asymptotically safe fixed point exists which
lies in the string landscape.

\section{Conclusions and outlook}
\label{sec:conclusions}
We have found indications that the weak gravity conjecture imposes
constraints on the microscopic parameter space of asymptotically safe
models.  This observation in itself is independent of the existence of
an embedding of the ASSM into string theory.

In a scenario with string theory as the fundamental theory of quantum
gravity, an intermediate asymptotically safe fixed point, see
Fig.~\ref{fig:ill_scales}, is expected to be subject to the weak
gravity conjecture. Moreover, such a scaling regime is a potential
candidate for the low-energy effective description emerging from
string theory. Our work, therefore, provides a first indication that
an asymptotically safe region might exist in the landscape.  We
highlight that the RG flow of an asymptotically safe scaling regime
could potentially connect a compactification of string theory on a
background with a negative \emph{microscopic} value of the
cosmological constant to infrared physics in dS space (i.e., with a
positive low-energy value of the cosmological constant).  We hasten to
add that further conditions beyond the weak-gravity conjecture should
be satisfied. Most importantly, we have not constructed a specific
choice of compactification, for which the coupling-values at $\bar{k}$
lie in the basin of attraction of the asymptotically safe fixed point,
and where $\bar{k}\gg k_{\rm tr}$. We simply point out that such a
construction could be possible.  In that region of the
string-theoretic landscape, the low-energy phenomenology of asymptotic
safety and string theory would essentially be indistinguishable. This
would, in particular, imply that first-principle calculations of
Standard Model couplings, which could be possible in asymptotic
safety, would also apply to string theory. On the other hand,
embedding asymptotic safety in a UV completion provided by string
theory places questions about unitarity in asymptotic safety
\cite{Becker:2017tcx,Arici:2017whq} in a different light. In a
string-embedding, asymptotic safety could even feature unstable
propagating modes. As long as their masses are at or above the string
scale, these instabilities simply constitute a signature for a more
fundamental UV completion and do not pose problems for the stability
of the theory. Accordingly, the class of fixed points that allows for the presented scenario might be larger than the class of fixed points that allows for fundamental asymptotic safety, where ghost modes should be absent.

There has been much discussion on the constraints on QFTs coming from
the requirement of a consistent coupling to quantum gravity. Most of
the discussion has been in the context of string theory - i.e., under
the assumption that quantum gravity corresponds to string
theory. Asymptotic safety also gives restrictions and has been
explored, e.g., in terms of its implications for chiral fermions
\cite{Eichhorn:2011pc}, a light Higgs \cite{Wetterich:2016uxm},
restrictions on the maximum number of matter fields
\cite{Dona:2013qba, Dona:2014pla,Meibohm:2015twa} and the allowed
interaction structures for matter
\cite{Eichhorn:2016esv,Eichhorn:2012va,Eichhorn:2017eht}. It is of
interest to understand to what extent such restrictions are compatible
(or in conflict with), the string theory restrictions, i.e., delineate
the boundaries and overlapping regions of the respective landscapes.

\acknowledgements{We thank A. Hebecker, K.~Stelle and L.~Freidel for
  discussions.  This work is supported by the Deutsche
  Forschungsgemeinschaft (DFG) under grant no.~Ei-1037/1. This
  research is also supported by the Danish National Research
  Foundation under grant DNRF:90. A.~H.~also acknowledges support by
  the German Academic Scholarship Foundation. A.~E.~is also supported
  by a visiting fellowship at the Perimeter Institute for Theoretical
  Physics. The work is part of and supported by the DFG Collaborative
  Research Centre SFB 1225 (ISOQUANT) as well as by the DFG under
  Germany's Excellence Strategy EXC-2181/1 - 390900948 (the Heidelberg
  Excellence Cluster STRUCTURES).}

\bibliography{refs}

\end{document}